%% file: DATE27-HLWalk.tex
\documentclass[10pt,conference]{IEEEtran}
\IEEEoverridecommandlockouts % show footnote comments
\input{setting-ieee}

\input{math_commands}

\graphicspath{{./figs/}{../}}

\usepackage[switch]{lineno}
\usepackage{tikz}
\usepackage{circuitikz}

\begin{document}

\title{\tool{HLWalk}: Modal-Split Random-Walk Thermal Analysis for Heterogeneous 2.5D/3D ICs}

\author{\IEEEauthorblockN{%
    Qibo Xue\textsuperscript{1*}, Shengnan Cai\textsuperscript{1*},
    Tianshu Hou\textsuperscript{1}, Tianxiang Zhu\textsuperscript{3},\\
    Zixiao Wang\textsuperscript{2\textdagger}, Yibo Lin\textsuperscript{3},
    Runsheng Wang\textsuperscript{3}, and Bei Yu\textsuperscript{1\textdagger}}
    \IEEEauthorblockA{%
    \textsuperscript{1}The Chinese University of Hong Kong\\
    \textsuperscript{2}The Hong Kong University of Science and Technology (Guangzhou)\\
    \textsuperscript{3}Peking University}}

\maketitle
\thispagestyle{plain}
\pagestyle{plain}

\input{docs/abs}

\input{docs/intro}
\input{docs/pre}
\input{docs/method}
\input{docs/exp}

\input{docs/conclusion}

\clearpage
\bibliographystyle{IEEEtran}
\bibliography{ref/Top-sim, ref/bib}

\end{document}

%% file: setting-ieee.tex
\usepackage{blkarray}                                      % to support matrix
\usepackage{graphicx}                                      % include pdf figures
\usepackage{amsmath}
\usepackage{amssymb}
\usepackage{amsfonts}
\usepackage{amsthm}
\usepackage[mathcal]{eucal}
\usepackage{mathrsfs}
\usepackage{booktabs}
\usepackage{enumerate}
\usepackage{multirow}
\usepackage[subrefformat=parens,farskip=0pt,justification=centering]{subfig}
\usepackage{color}
\usepackage{cite}                                          % more citations in one bracket
\usepackage{comment}                                       % use comment
\usepackage{soul}                                          % use highlight command \hl{}
\soulregister\cite7
\soulregister\ref7
\soulregister\pageref7
\usepackage{etoolbox}                                      % commands \newtoggle, \toggletrue, \iftoggle
\usepackage{url}
\usepackage{nth}                                           % nth command
\usepackage{bm}                                            % bm command
\usepackage{courier}
\usepackage{balance}
\usepackage{threeparttable}
\usepackage{xcolor,colortbl}
\usepackage{footnote}
\usepackage{listings}
\usepackage{setspace}                                      % setstretch{}
\usepackage[inline]{enumitem}
\usepackage{verbatim}
\usepackage[bookmarks=false]{hyperref}
\hypersetup{
    colorlinks = true,
    citecolor  = blue,
    linkcolor  = blue,
    urlcolor   = blue,
}
\usepackage{tikz}
\usetikzlibrary{patterns,snakes}
\usetikzlibrary{positioning,calc,fit,decorations.pathmorphing,shapes.geometric, shapes.gates.logic.US, calc}
\usetikzlibrary{arrows,arrows.meta,decorations.markings,shapes,shapes.arrows}
\usetikzlibrary{decorations,decorations.pathreplacing}
\usetikzlibrary{backgrounds}
\usepackage{filecontents}                                  % support to pgfplots
\usepackage{pgfplots}
\usepackage{pgfplotstable}
\usepgfplotslibrary{groupplots}
\usepackage{scalefnt}
\pgfplotsset{compat=newest}
\usepackage{caption}
\usepackage{pifont}                                        % ding commend
\usepackage{cleveref}
\Crefformat{figure}{Fig.~#2#1#3}                           % "Fig.", instead of "Figure"
\Crefname{subfigure}{Fig.}{Figs.}
\Crefname{figure}{Fig.}{Figs.}
\Crefformat{table}{TABLE~#2#1#3}                           % "TABLE", instead of "Table"
\usepackage[figuresright]{rotating}
\usepackage{algorithm}
\iftrue
\usepackage{algpseudocode}                                 % new algorithm package
\algrenewcommand\textproc{\texttt}
\makeatletter
\let\OldStatex\Statex
\renewcommand{\Statex}[1][3]{%
  \setlength\@tempdima{\algorithmicindent}%
  \OldStatex\hskip\dimexpr#1\@tempdima\relax
}
\makeatother
\else
\usepackage{algorithmic}
\fi

\definecolor{CUHKorange}{RGB}{244,106,18} %F47012
\definecolor{CUHKblue}{RGB}{0,111,190}    %006FBE
\definecolor{CUHKgreen}{RGB}{0,127,128}   %007F80
\definecolor{CUHKred}{RGB}{228,46,36}     %E42E24
\definecolor{CUHKyellow}{RGB}{198,148,34} %C69422
\definecolor{CUHKdark}{RGB}{114,44,114}   %722C72
\definecolor{CUHKmiddle}{RGB}{144,44,144} %902C90
\definecolor{CUHKlight}{RGB}{167,44,167} 
\definecolor{CUHKpurple}{RGB}{117,15,109}
\definecolor{CUHKgold}{RGB}{221,163,0}
\definecolor{CUHKribbon}{RGB}{244,223,176}
\definecolor{CUHKblack}{RGB}{34,24,21}

\newcommand{\calO}{\mathcal{O}}

\newcommand{\tool}[1]{$\mathsf{#1}$}
\renewcommand{\bf}[1]{\textbf{#1}}
\renewcommand{\tt}[1]{\texttt{#1}}
\renewcommand{\vec}[1]{\boldsymbol{#1}}    % re-define vec command

\newcommand{\minisection}[1]{\vspace{.05in}\noindent{\textbf{#1}}.}

\usepackage{tcolorbox}
\tcbuselibrary{skins,breakable}
    {\endtcolorbox}
    {\endtcolorbox}

\usepackage[margin=1in]{geometry}
\crefname{mytheorem}{Theorem}{Theorems}
\crefname{mylemma}{Lemma}{Lemmas}
\crefname{myclaim}{Claim}{Claims}
\crefname{myproperty}{Property}{Properties}
\crefname{mycorollary}{Corollary}{Corollaries}

\newcommand{\tabincell}[2]{\begin{tabular}{@{}#1@{}}#2\end{tabular}}

\RequirePackage[normalem]{ulem} %DIF PREAMBLE
\RequirePackage{color}\definecolor{RED}{rgb}{1,0,0}\definecolor{BLUE}{rgb}{0,0,1} %DIF PREAMBLE

%% file: math_commands.tex
\usepackage{amsmath,amsfonts,bm}

\def\eqref#1{equation~\ref{#1}}
\def\ceil#1{\lceil #1 \rceil}

\def\1{\bm{1}}

\def\eps{{\epsilon}}

\def\rvx{{\mathbf{x}}}
\def\rvy{{\mathbf{y}}}

\def\vtheta{{\bm{\theta}}}
\def\va{{\bm{a}}}
\def\vb{{\bm{b}}}
\def\vc{{\bm{c}}}

\def\vg{{\bm{g}}}

\def\vn{{\bm{n}}}

\def\vq{{\bm{q}}}
\def\vr{{\bm{r}}}

\def\vv{{\bm{v}}}

\def\vx{{\bm{x}}}
\def\vy{{\bm{y}}}

\def\evb{{b}}

\def\evg{{g}}

\def\mA{{\bm{A}}}

\def\mI{{\bm{I}}}

\def\mK{{\bm{K}}}

\def\mV{{\bm{V}}}

\DeclareMathAlphabet{\mathsfit}{\encodingdefault}{\sfdefault}{m}{sl}
\SetMathAlphabet{\mathsfit}{bold}{\encodingdefault}{\sfdefault}{bx}{n}

\def\emK{{K}}

\newcommand{\E}{\mathbb{E}}

\newcommand{\R}{\mathbb{R}}

%% file: docs/abs.tex
\begin{abstract}
Efficient thermal analysis is essential for accelerating the design
of 2.5D/3D ICs. However, existing random-walk thermal solvers can
suffer from boundary discretization errors and high computational
costs when resolving fine-scale structures such as through-silicon
vias (TSVs). To address these challenges, we propose \tool{HLWalk},
an efficient random-walk thermal solver for heterogeneous
2.5D/3D ICs. We introduce local boundary integral reconstruction
to eliminate discretization errors in local-time-based boundary treatments.
To accelerate boundary preprocessing, we employ modal splitting:
low-order modes are computed deterministically, while Monte Carlo
sampling estimates only the residual. We further develop
scale-aware random-walk transport to efficiently resolve fine-scale
structures without requiring fine discretization throughout the domain. Experiments on 2.5D/3D IC benchmarks show that \tool{HLWalk} achieves remarkable wall-clock speedup and better accuracy across settings.
\end{abstract}

%% file: docs/intro.tex
\section{Introduction}

Heterogeneous integration (HI) technologies enable continued system-level scaling~\cite{srcmapt}. However, increasing integration density and power density exacerbate hotspots and temperature rise, significantly degrading chip performance, reliability, and lifetime~\cite{Liu2014:TVLSI,puttaswamy2006thermal}. Meanwhile, design-space exploration often requires hundreds or thousands of repeated heat-conduction partial differential equation (PDE) solves, imposing substantial computational overhead~\cite{TAP-2.5D,Wang2026:TCAD,sultan2020:DAC}. Therefore, effective thermal estimation is critical for the design and optimization of 2.5D/3D ICs~\cite{3D-ICE,PACT,liu2023deepoheat}.

Deterministic methods, such as the Finite Element Method (FEM), can provide high-fidelity temperature distributions~\cite{Sultan2019:CS,MTA_1}. However, they require constructing a numerical mesh over the entire
thermal domain and solving the resulting large linear systems, making repeated thermal analysis expensive, especially for geometrically complex structures requiring high spatial resolution.

\begin{figure}[!t]
    \centering
    \includegraphics[width=0.98\columnwidth]{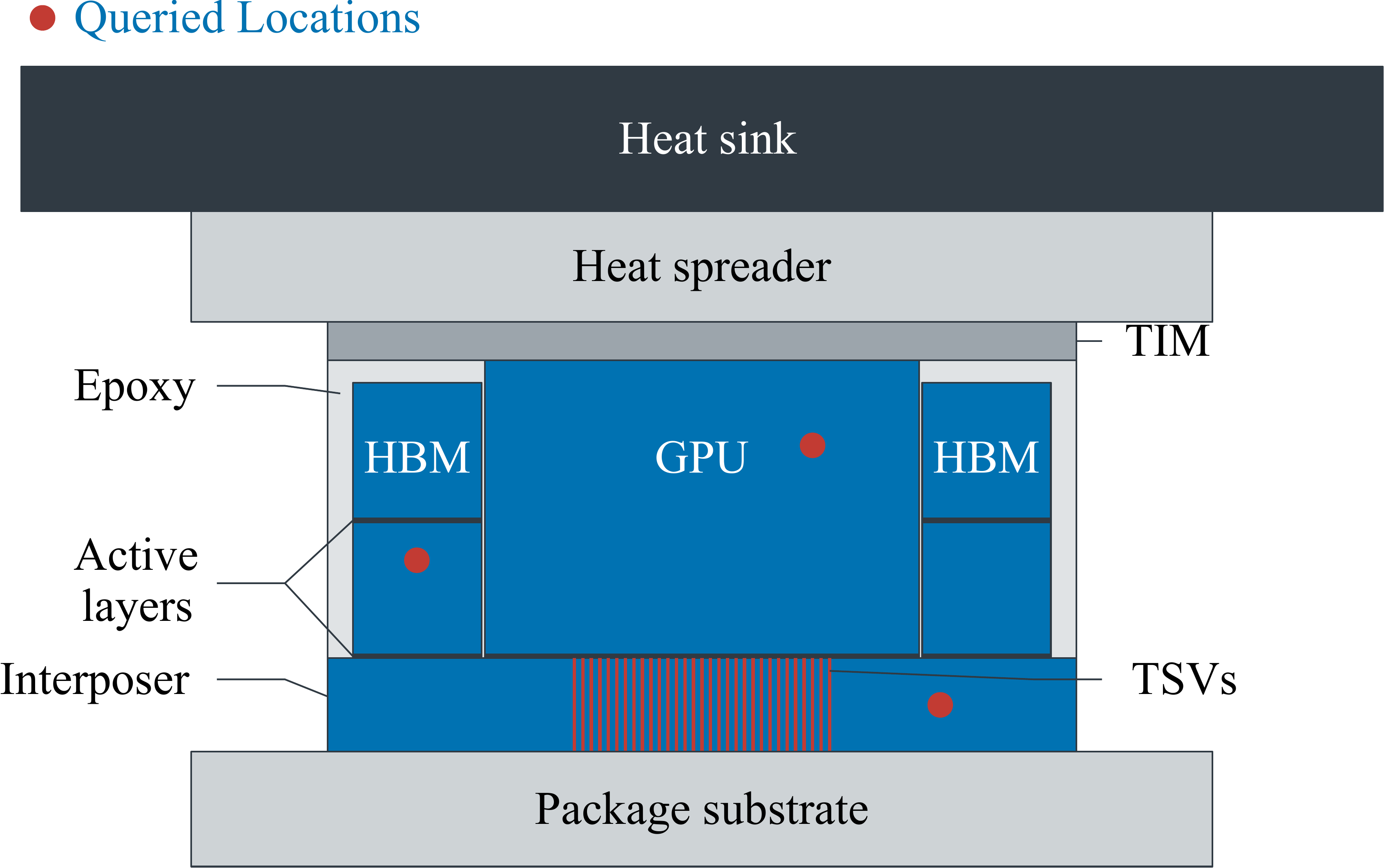}
    \caption{An illustration of a 2.5D IC with TSVs. In
certain scenarios, only the temperatures at a few queried
locations are needed.}
    \label{fig:queried_locations}
\end{figure}

Nevertheless, full-chip temperature distributions are not always necessary, as chip design often focuses more on hotspots and some critical locations (\Cref{fig:queried_locations}). Random-walk methods are well suited for localized thermal estimation because of their inherent parallelism and have shown success in thermal analysis and other EDA applications with Dirichlet boundary conditions~\cite{Iverson2000:SISPAD,Qian2005:TCAD,Yu2013:TCAD,Marios2023:TCAD}. Subsequent work introduced hybrid random-walk schemes to handle adiabatic Neumann and convective Robin boundaries~\cite{Liang2014:ASPDAC}, which were later enhanced with Feynman--Kac formulations for improved accuracy~\cite{Yang2022:ICCAD,Dong2025:TCAD}. However, existing local-time-based methods suffer from severe boundary discretization errors ~\cite{Zhou2017:CMS,Zhou2016:JSC}. Boundary integral equation (BIE)-based methods alleviate these issues, but they introduce substantial boundary precomputation overhead and are directly applicable to homogeneous source-free regions~\cite{Sabelfeld1982,Sabelfeld1991,SabelfeldSimonov1994}. Moreover, existing random-walk methods struggle to maintain both accuracy and efficiency in heterogeneous 2.5D/3D ICs with large geometric scale disparities.

To overcome these limitations, we propose HLWalk (``HL'' refers to high- and low-mode decomposition), an efficient random-walk solver for steady-state thermal analysis of 2.5D/3D ICs. HLWalk eliminates boundary discretization error through local boundary reconstruction. To reduce boundary-processing overhead, it separates deterministic low-order components from the stochastic residual and applies small Monte Carlo boundary caching only to the unresolved part, with deterministic correction of its low-frequency components. For heterogeneous 2.5D/3D structures with large geometric scale disparities, HLWalk employs scale-aware random-walk transport, using Walk-on-Spheres (WOS) in large homogeneous regions and locally refined, conductivity-aware Walk-on-Grids (WOG) around fine-scale features such as TSVs. The main contributions of this work are summarized as follows:

\begin{itemize}
    \item An efficient thermal solver that eliminates local-time boundary discretization error.
    
    \item A modal-split acceleration scheme which significantly reduces processing cost while enhancing accuracy.
    
    \item A scale-aware WOS/WOG strategy for heterogeneous 2.5D/3D ICs with large
geometric scale disparities.
    
    \item Comprehensive experimental evaluation on representative
    2.5D/3D IC benchmarks, demonstrating the accuracy
    and efficiency of HLWalk.
\end{itemize}

%% file: docs/pre.tex
\section{Preliminaries}
\label{sec:preliminaries}

\subsection{Problem Formulation}
\label{subsec:problem_formulation}

We model a heterogeneous 2.5D/3D IC as a three-dimensional thermal
domain $\Omega\subset\R^{3}$ composed of multiple materials. Under
steady-state conditions, the temperature field $T(\vx)$ is governed by
\begin{equation}
-\nabla\cdot
\left(
k(\vx)\nabla T(\vx)
\right)
=
p(\vx),
\qquad
\vx\in\Omega,
\label{eq:thermal_problem}
\end{equation}
where $k(\vx)$ denotes the spatially varying thermal conductivity and
$p(\vx)$ denotes the volumetric heat-generation density.

The external boundary is partitioned into
$\Gamma_{\mathrm{D}}$, $\Gamma_{\mathrm{N}}$, and
$\Gamma_{\mathrm{R}}$, on which Dirichlet, Neumann, and Robin
conditions are imposed, respectively:
\begin{align}
T(\vx)
&=
T_{\mathrm{D}}(\vx),
&&
\vx\in\Gamma_{\mathrm{D}},
\label{eq:bc_dirichlet}
\\
-k(\vx)\partial_{\vn}T(\vx)
&=
q_{\mathrm{N}}(\vx),
&&
\vx\in\Gamma_{\mathrm{N}},
\label{eq:bc_neumann}
\\
-k(\vx)\partial_{\vn}T(\vx)
&=
h(\vx)
\left[
T(\vx)-T_{\mathrm{amb}}
\right],
&&
\vx\in\Gamma_{\mathrm{R}},
\label{eq:bc_robin}
\end{align}
where $\partial_{\vn}T=\nabla T\cdot\vn$ and $\vn$ is the outward
unit normal. Here, $T_{\mathrm{D}}(\vx)$ is the prescribed boundary
temperature, $q_{\mathrm{N}}(\vx)$ is the normal heat-flux density,
$h(\vx)$ is the convection coefficient, and $T_{\mathrm{amb}}$ is the
ambient temperature. An adiabatic surface is represented by
$q_{\mathrm{N}}=0$.

Heterogeneous materials further introduce internal interfaces. Across each material interface, temperature and normal heat flux remain continuous:
\begin{equation}
T^{-}=T^{+},
\qquad
k^{-}\partial_{\vn}T^{-}
=
k^{+}\partial_{\vn}T^{+}.
\label{eq:interface_condition}
\end{equation}

Rather than recovering the complete temperature field, we consider a
set of query locations $\{\vx_q\}_{q=1}^{N_q}$ and seek to efficiently
estimate
\begin{equation}
T(\vx_q),
\qquad
q=1,\ldots,N_q,
\label{eq:query_objective}
\end{equation}
for heterogeneous 2.5D/3D ICs characterized by large geometric scale
disparities.

\subsection{Random Walk for Thermal Simulation}
\label{subsec:random_walk}

Random-walk thermal solvers typically employ two complementary
transport schemes, WOG and WOS,
for heterogeneous and homogeneous regions, respectively~\cite{Liang2014:ASPDAC,Yang2022:ICCAD}.

\minisection{Walk-on-Grids}
WOG discretizes the thermal domain into voxels and models heat diffusion through transitions between neighboring cells. For an orthogonal grid, the effective thermal conductance is
\begin{equation}
G_{ij}
=
\frac{A_{ij}}
{\displaystyle
\frac{d_i}{k_i}
+
\frac{d_j}{k_j}},
\label{eq:wog_conductance}
\end{equation}
where $A_{ij}$ is the shared interface area, $d_i$ and $d_j$ are the
distances to the interface, and $k_i$ and $k_j$ are the corresponding
thermal conductivities.
Thus, the transition probability between adjacent cells $i$ and $j$ can be denoted as:
\begin{equation}
P_{ij}
=
\frac{G_{ij}}
{\displaystyle\sum_{m\in\mathcal{N}(i)}G_{im}},
\label{eq:wog_probability}
\end{equation}
where $\mathcal{N}(i)$ denotes the neighboring cells of $i$. 

\minisection{Walk-on-Spheres}
In homogeneous source-free regions, the governing equation reduces to
the Laplace equation. By the mean-value property,
\begin{equation}
T(\vx)
=
\frac{1}{4\pi r^{2}}
\int_{\partial \mathcal{B}(\vx,r)}
T(\vy)\,dS_{\vy}
=
\E\!\left[T(\rvy)\right],
\label{eq:wos_mean_value}
\end{equation}
where $\mathcal{B}(\vx,r)$ is a sphere centered at $\vx$ and $\rvy$ is
uniformly sampled on its surface. WOS directly samples the temperature
on the sphere boundary, allowing large spatial jumps without resolving
the intermediate trajectory~\cite{Sawhney2023:TOG,Miller2024:TOG}.

\begin{figure}[!t]
    \centering
    \includegraphics[width=0.828\columnwidth]
    {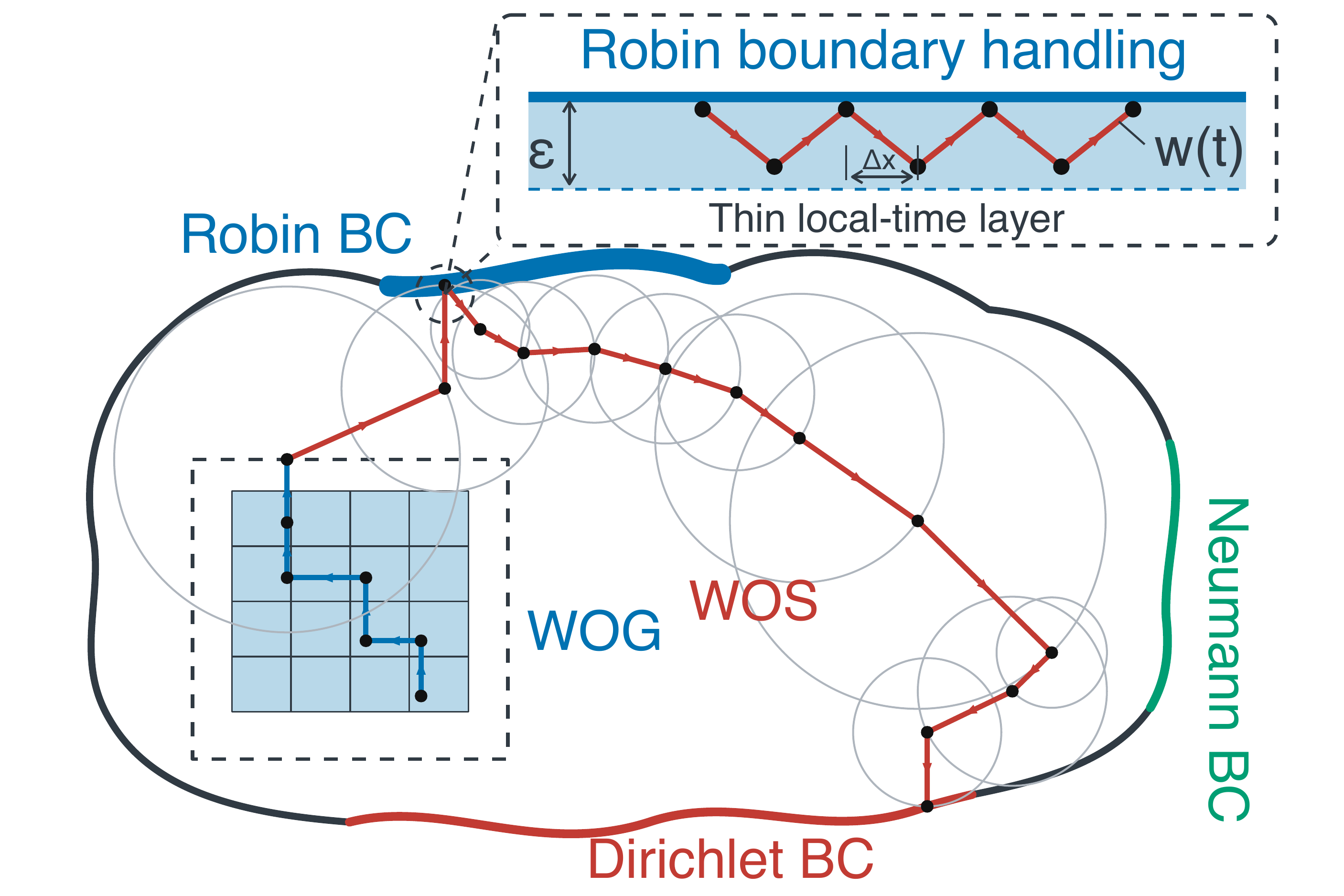}
    \caption{An illustration of local-time-based random-walk thermal transport with WOG and WOS.}
    \label{fig:rw_overview}
\end{figure}

As \Cref{fig:rw_overview} shows, WOG is therefore suited to regions containing heat sources, material
interfaces, or fine geometric features by capturing local material and geometric variations, while WOS provides efficient
transport in homogeneous source-free regions.  

\subsection{Challenges of Existing Random-Walk Thermal Solvers}
\label{subsec:rw_challenges}

Despite their efficiency in localized temperature estimation, existing
random-walk thermal solvers face two major challenges in heterogeneous
2.5D/3D ICs: severe boundary discretization error and large geometric scale
disparities.

\minisection{Boundary discretization}
For Neumann and Robin boundaries, Feynman--Kac-based methods describe
boundary interactions through reflecting Brownian motion and its
boundary local time $L(t)$~\cite{oksendal2003stochastic,Zhou2017:CMS,Zhou2016:JSC}. For a boundary neighborhood
$\mathcal{D}_{\eps}$ of width $\eps$,
\begin{equation}
L(t)
=
\lim_{\eps\rightarrow 0}
\frac{1}{\eps}
\int_{0}^{t}
\1_{\mathcal{D}_{\eps}}(\rvx_s)\,ds .
\label{eq:local_time}
\end{equation}
Since this limiting process cannot be evaluated directly, numerical
solvers approximate it using a finite boundary strip and random-walk
step size, e.g.,
\begin{equation}
\Delta L
\approx
\delta n
\frac{(\Delta x)^2}{3\eps},
\label{eq:local_time_discrete}
\end{equation}
where $\Delta x$ is the near-boundary step size and $\delta n$ denotes
the accumulated boundary interactions~\cite{Yang2022:ICCAD,Dong2025:TCAD}.

For Robin boundaries, the local time further determines the
Feynman--Kac attenuation weight~\cite{Zhou2016:JSC},
\begin{equation}
w(t)
=
\exp\left(
-\frac{1}{2}
\int_{0}^{t}
c(\rvx_s)\,dL(s)
\right),
\label{eq:fk_weight}
\end{equation}
where $c(\vx)$ is determined by the Robin boundary condition.
Therefore, errors in the numerical approximation of $L(t)$ perturb the accumulated path weight and each path may interact with the boundary many times within the local‑time layer prior to hitting the stop threshold, so such discretization errors will influence the temperature prediction significantly. Moreover, the numerical approximation of $L(t)$ makes the solution sensitive to the choice of discretization parameters. \Cref{fig:rw_overview} shows the local-time-based boundary process.

\minisection{Large geometric scale disparities}
Heterogeneous 2.5D/3D ICs combine large structures such as dies and
interposers with much finer structures like TSVs~\cite{Iverson2000:SISPAD}. Let $L_{\alpha}$ and $\ell_{\alpha}$ denote the large and
small characteristic scales along direction $\alpha$.
Their scale disparity is characterized by
\begin{equation}
\chi_{\alpha}
=
\frac{L_{\alpha}}{\ell_{\alpha}},
\qquad
\alpha\in\{x,y,z\},
\label{eq:scale_ratio}
\end{equation}
where $\chi_{\alpha}\gg1$ for strongly multiscale structures.

If the entire domain is discretized according to the smallest local
scale, the number of cells grows approximately as
\begin{equation}
N_{\mathrm{cell}}
=
O\left(
\chi_x\chi_y\chi_z
\right).
\label{eq:grid_scale_complexity}
\end{equation}
For random-walk methods, uniformly fine grids are costly, while coarse grids may miss fine-scale features, making multiscale thermal analysis challenging.

%% file: docs/method.tex
\section{HLWalk Method}
\label{sec:method}
In this section, we present our proposed random-walk solver, \tool{HLWalk}. First, \Cref{subsec:bie} presents the combination of the BIE method with hybrid random walk for local boundary reconstruction, which accounts for the effect of internal heat sources while avoiding discretization errors. Building on this reconstruction, \Cref{subsec:modal_split} introduces a modal-split scheme that significantly reduces the boundary processing cost with accuracy enhancement. \Cref{subsec:scale_adaptive} addresses 2.5D/3D ICs featuring complex material distributions and disparate geometric scales by adopting scale-aware WOS/WOG transport, adjusting WOG spatial resolution based on geometry and materials.

\subsection{Boundary Reconstruction for Chip Thermal Analysis}
\label{subsec:bie}

Define the temperature rise relative to the ambient temperature as
\begin{equation}
\theta(\vx)
=
T(\vx)-T_{\mathrm{amb}}.
\label{eq:theta_definition}
\end{equation}

For the $i$-th Robin heat-dissipation region, the boundary condition is
written as
\begin{equation}
-k_i\frac{\partial \theta}{\partial n}
=
h_i\theta,
\label{eq:robin_theta}
\end{equation}
where $k_i$ and $h_i$ denote the local thermal conductivity and
convective heat-transfer coefficient, respectively.

For a local boundary patch $S_i$, an auxiliary hemisphere $\Gamma_i$
is introduced. Substituting the Robin boundary condition into the
local BIE yields, for $\vx\in S_i$,
\begin{equation}
\begin{aligned}
\frac{1}{2}\theta(\vx)
&+
\frac{h_i}{k_i}
\int_{S_i}
G_1(\vx,\vy)
\theta(\vy)\,dS_{\vy} \\
&=
\int_{\Gamma_i}
P_i(\vx,\vy)
\theta(\vy)\,dS_{\vy}.
\end{aligned}
\label{eq:local_bie}
\end{equation}
Here, $G_1$ denotes the local spherical Green's function, and
\begin{equation}
P_i(\vx,\vy)
=
-
\frac{\partial G_1(\vx,\vy)}
{\partial n_{\vy}}
\label{eq:bie_kernel}
\end{equation}
is the corresponding boundary kernel.
\Cref{eq:local_bie} relates the Robin-boundary temperature rise to the auxiliary-hemisphere temperature through local conduction and convection parameters.

To account for internal heat sources and heterogeneous materials during boundary reconstruction, \tool{HLWalk} couples auxiliary-surface temperatures with scale-aware WOG/WOS thermal transport. The response at the center of the $i$-th boundary patch is written as
\begin{equation}
\evb_i
=
\va_i^{\mathsf T}
\vtheta_{\Gamma_i},
\label{eq:local_response}
\end{equation}
where $\va_i$ contains the response weights determined by
the local geometry, material parameters, and Robin boundary.

For a sampling location $\vy$ on the auxiliary hemisphere,
\tool{HLWalk} represents the temperature as
\begin{equation}
\theta(\vy)
=
\E_{\vy}
\left[
R(\omega)
+
b(\vec{Y}_\omega)
\right],
\label{eq:hemisphere_transport}
\end{equation}
where \(R(\omega)\) is the accumulated heat-source contribution along path \(\omega\), and \(\vec{Y}_\omega\) is its terminal location on the heat-dissipation boundary. Scale-aware WOG captures contributions from heat-source and heterogeneous multiscale regions, while WOS is used in homogeneous source-free regions.

Since the temperatures on Robin boundaries are unknown, different
boundary patches are coupled through random paths:
\begin{equation}
\evb_i
=
\evg_i
+
\sum_j \emK_{ij}\evb_j,
\label{eq:boundary_coupling}
\end{equation}
which leads to the global boundary system
\begin{equation}
(\mI-\mK)\vb
=
\vg.
\label{eq:boundary_system}
\end{equation}
Here, \(\vg\) denotes the heat-source contribution and \(\mK\) the stochastic boundary coupling. The boundary temperature rises are obtained by solving \Cref{eq:boundary_system} with preconditioned GMRES.

\subsection{Modal-Split Acceleration}
\label{subsec:modal_split}

\tool{HLWalk} reduces stochastic computation by separating smooth
low-order components from the residual thermal field, while boundary preprocessing is further
accelerated using sparse sampling with deterministic low-mode correction.

\minisection{Residual Decomposition}
Chip temperature fields typically contain smooth, large-scale spatial
components that can be efficiently represented by low-order modes.
To avoid repeatedly estimating these components through stochastic
sampling, the thermal problem is written as
\begin{equation}
\mA\vtheta
=
\vq.
\label{eq:full_system}
\end{equation}
\tool{HLWalk} constructs a deterministic control field
using a low-order cosine Galerkin space $\mV_0$:
\begin{equation}
\left(
\mV_0^{\mathsf T}\mA\mV_0
\right)\vc_0
=
\mV_0^{\mathsf T}\vq,
\qquad
\vtheta_0
=
\mV_0\vc_0.
\label{eq:control_field}
\end{equation}
To remove the smooth layer-wise background at minimal deterministic
cost, \(\mV_0\) is formed by the zero-order in-plane cosine mode of
each layer. Projecting \(\mA\) onto this space preserves inter-layer
thermal coupling while extracting the dominant large-scale temperature
component.

\begin{figure}[!t]
    \centering
    \includegraphics[width=0.98\columnwidth]{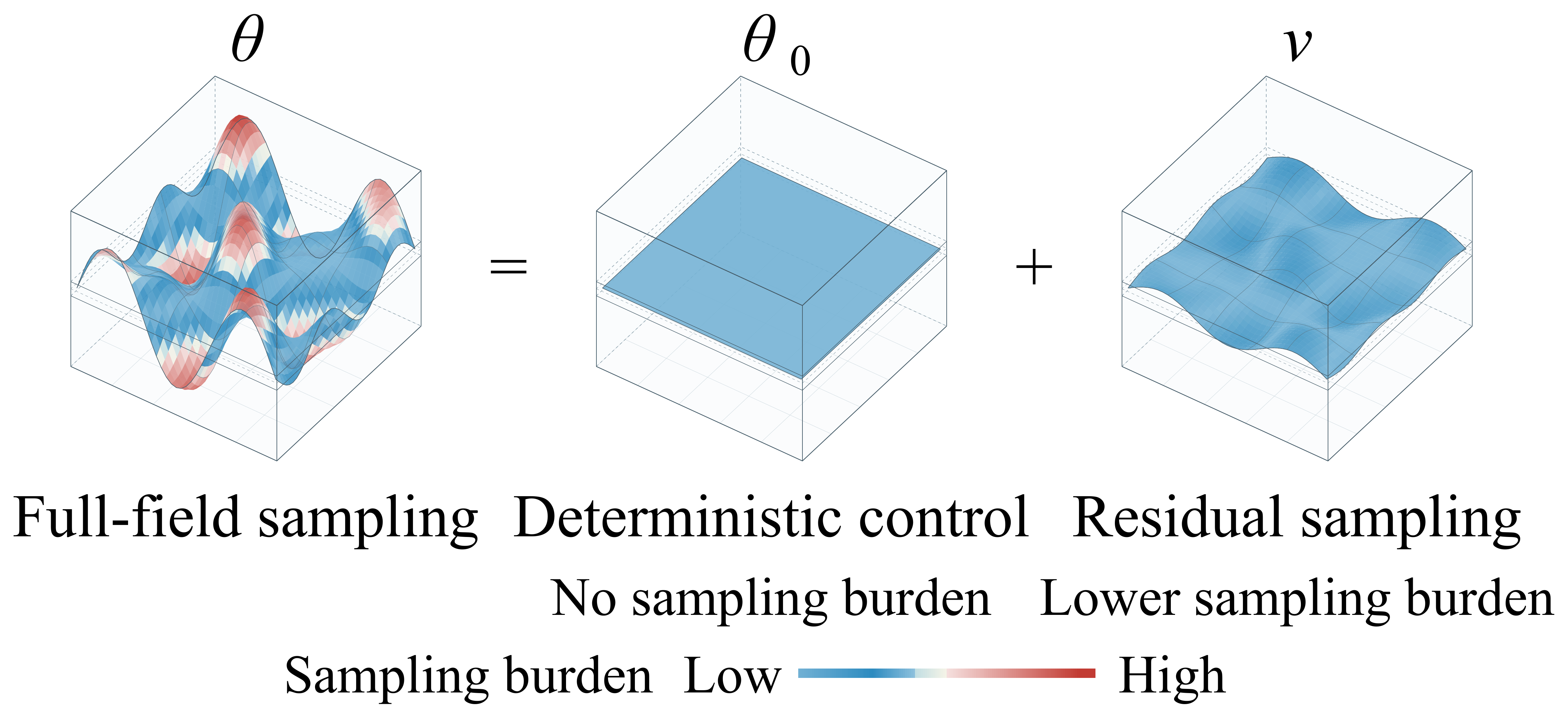}
    \caption{An illustration of residual decomposition in \tool{HLWalk}.}
    \label{fig:hlwalk_residual}
\end{figure}

As shown in \Cref{fig:hlwalk_residual}, the temperature field is decomposed as
\begin{equation}
\vtheta
=
\vtheta_0
+
\vv,
\label{eq:field_decomposition}
\end{equation}
where the residual field satisfies
\begin{equation}
\mA\vv
=
\vr,
\qquad
\vr
=
\vq
-
\mA\vtheta_0.
\label{eq:residual_system}
\end{equation}
The stochastic solver is applied only to the residual problem.
The boundary field is correspondingly decomposed as
\begin{equation}
\vb
=
\vb_0
+
\vb_r,
\qquad
(\mI-\mK)\vb_r
=
\vg_r,
\label{eq:residual_boundary_system}
\end{equation}
where \(\vb_0\) is the boundary value of the control field and
\(\vg_r = \vg-(\mI-\mK)\vb_0\) is the residual boundary source term.
By removing the smooth large-scale component before stochastic
sampling, the residual problem can be estimated with fewer random-walk
paths.

\minisection{Sparse Boundary Sampling}
At the boundary processing stage, Monte Carlo sampling over the auxiliary hemispheres accounts for most of the cost. With $N_b$ boundary
patches, $N_s$ sampling points per hemisphere, $N_c$ paths per sampling
point, and average path length $\bar L$, the cost is approximately
\begin{equation}
\mathcal C_{\mathrm{cache}}
= 
\calO
\left(
N_bN_sN_c\bar L
\right).
\label{eq:cache_complexity}
\end{equation}

After residual decomposition, \tool{HLWalk} constructs the boundary
cache for the residual problem. For a hemispherical sampling point
$\vy$, its residual temperature is estimated by
\begin{equation}
\widehat v(\vy)
=
\frac{1}{N_c}
\sum_{\ell=1}^{N_c}
\left[
R_r(\omega_\ell)
+
b_r(\vec{Y}_\ell)
\right],
\label{eq:residual_mc}
\end{equation}
where $R_r(\omega_\ell)$ is the accumulated residual reward and
$\vec{Y}_\ell$ is the terminal boundary location.

Each trajectory stores only its terminal boundary state and residual
reward. Terminal temperatures are interpolated from neighboring patches, and the samples are assembled into
\begin{equation}
(\mI-\widehat{\mK})
\widehat{\vb}_r
=
\widehat{\vg}_r.
\label{eq:approx_boundary_system}
\end{equation}
where $\widehat{\vb}_r$ denotes the Monte Carlo estimate of the residual
boundary field. Reducing $N_c$ lowers preprocessing and storage costs
but increases sampling error, motivating deterministic low-frequency correction.

\minisection{Low-Mode Enhancement}
To improve the boundary accuracy under a small cache size, \tool{HLWalk} employs
low-mode enhancement by further computing a deterministic approximation
in a richer cosine Galerkin space $\mV_L$:
\begin{equation}
\left(
\mV_L^{\mathsf T}\mA\mV_L
\right)\vc_L
=
\mV_L^{\mathsf T}\vq,
\qquad
\vtheta_L
=
\mV_L\vc_L.
\label{eq:low_mode_system}
\end{equation}
Compared with $\mV_0$, $\mV_L$ retains multiple low-order cosine modes, better representing smooth large-scale variations for boundary correction.

Let $\vb_L$ be $\vtheta_L$ on the boundary, with residual
\begin{equation}
\vb_L^r
=
\vb_L
-
\vb_0.
\label{eq:low_mode_residual}
\end{equation}

\begin{figure}[tb!]
    \centering
    \includegraphics[
        height=4.3cm,
        keepaspectratio
    ]{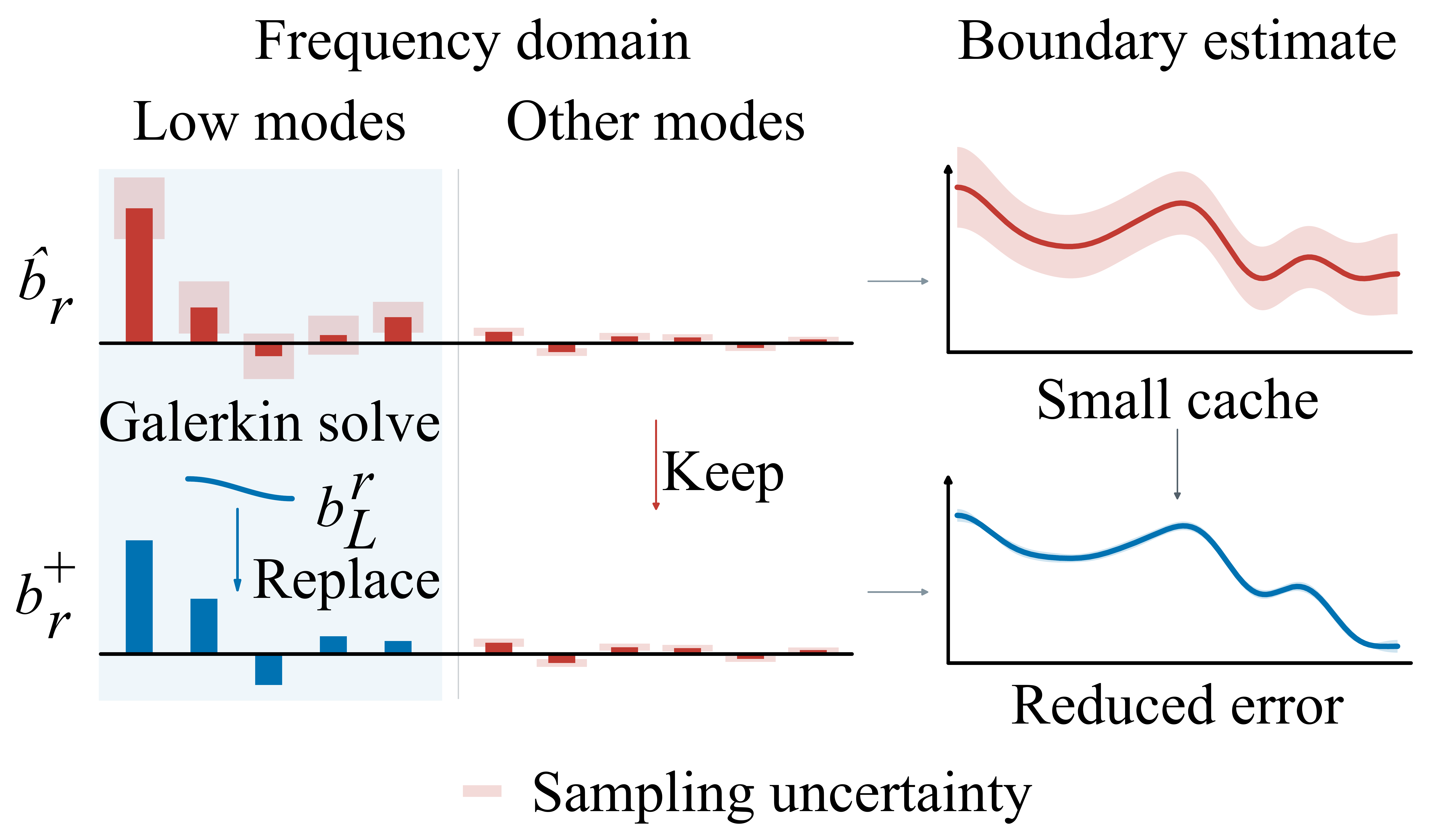}
    \caption{An illustration of low-mode enhancement in \tool{HLWalk}.}
    \label{fig:low_mode_enhancement}
\end{figure}

Then the boundary correction is performed in the spatial-frequency domain.
Discrete cosine transform (DCT) is applied
to each heat-dissipation boundary face, and selected
low-frequency coefficients of the stochastic residual boundary are
replaced by those from $\vb_L^r$. The residual boundary temperature rise is then combined from the Monte Carlo and low-mode components:
\begin{equation}
\vb_r^+
=
(\mI-\Pi_B)
\widehat{\vb}_r
+
\Pi_B
\vb_L^r,
\label{eq:low_mode_correction}
\end{equation}
where $\Pi_B$ projects onto selected low-frequency boundary DCT modes,
with the remaining components retained from Monte Carlo sampling. This
corrects the smooth error from small-cache sampling while
preserving its efficiency. Therefore, the complete boundary temperature rise is
\begin{equation}
\vb
=
\vb_0
+
\vb_r^+.
\label{eq:final_boundary}
\end{equation}
For a query location $\vx_q$, the final temperature estimate is
\begin{equation}
\begin{aligned}
\widehat T(\vx_q)
&=
T_{\mathrm{amb}}
+
\theta_0(\vx_q) \\
&\quad+
\frac{1}{N}
\sum_{\ell=1}^{N}
\left[
R_r(\omega_{q\ell})
+
b_r^+(\vec{Y}_{q\ell})
\right].
\end{aligned}
\label{eq:query_temperature}
\end{equation}
where $N$ is the number of random-walk paths for query $\vx_q$.

\Cref{fig:low_mode_enhancement} shows how to reduce small-cache error through low-mode enhancement.

\subsection{Scale-Aware Transport for Multiscale 2.5D/3D ICs}
\label{subsec:scale_adaptive}

\begin{figure}[!t]
    \centering
    \includegraphics[
        height=4.3cm,
        keepaspectratio
    ]{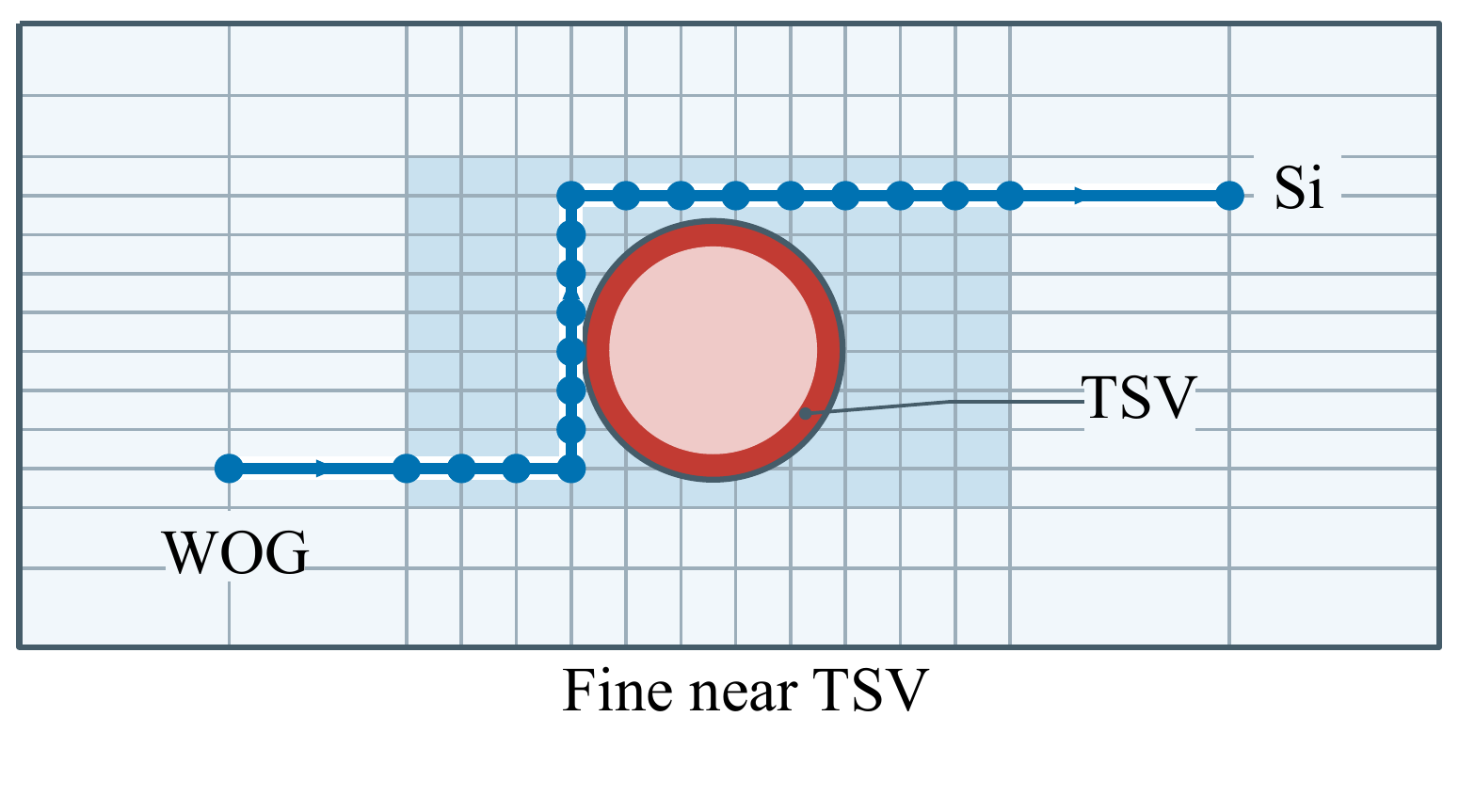}
    \caption{An illustration of scale-aware WOG near a TSV in \tool{HLWalk}.}
    \label{fig:hlwalk_wog}
\end{figure}

Advanced 2.5D/3D ICs span large package structures and fine features such as TSVs, creating strong geometric and material heterogeneity. Uniform fine grids are costly, while coarse grids miss local details.

As \Cref{fig:hlwalk_wog} shows, \tool{HLWalk} constructs a nonuniform orthogonal grid according to
local geometric scales. 
Boundaries of fine-scale structures such as TSVs and material interfaces are inserted as grid
breakpoints. Consequently, the local cell dimensions $(\Delta x_i,\Delta y_j,\Delta z_k)$
vary according to the local structural scale. Smaller cells are confined
to regions requiring explicit resolution of fine geometry or material
transitions, while larger cells are retained elsewhere.

For fine-scale structures with large scale disparities, such as the
$v$-th TSV, let $(x_v,y_v)$ denote its lateral center, $r_v$ its radius,
and $\delta_v$ the target local spatial resolution. Its lateral
diameter is divided into at least
\begin{equation}
N_v
=
\ceil{
\frac{2r_v}{\delta_v}
}
\label{eq:tsv_resolution}
\end{equation}
intervals, allowing the local resolution to follow the characteristic structural scale rather than a uniform global grid.

Each cell inherits local material properties, preserving fine features such as TSVs, while WOG transitions follow \Cref{eq:wog_conductance}.

WOS handles homogeneous source-free regions, while scale-aware WOG resolves heterogeneous and heat-source regions, capturing fine features such as TSVs without globally using the finest resolution.

%% file: docs/exp.tex
\begin{figure*}[!t]
    \centering
    \captionsetup[subfloat]{font=normalsize}

    \begin{minipage}[t]{0.265\textwidth}
        \vspace{0pt}
        \centering
        \subfloat[Cube]{
            \includegraphics[width=0.97\linewidth]
            {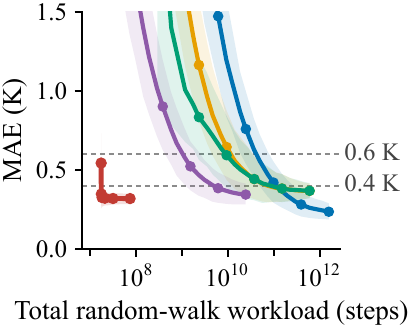}
        }
    \end{minipage}%
    \hfill
    \begin{minipage}[t]{0.265\textwidth}
        \vspace{0pt}
        \centering
        \subfloat[3DIC]{
            \includegraphics[width=0.97\linewidth]
            {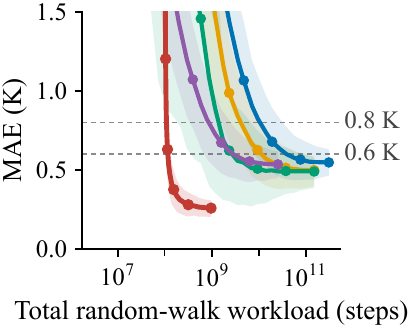}
        }
    \end{minipage}%
    \hfill
    \begin{minipage}[t]{0.265\textwidth}
        \vspace{0pt}
        \centering
        \subfloat[GPU+HBM]{
            \includegraphics[width=0.97\linewidth]
            {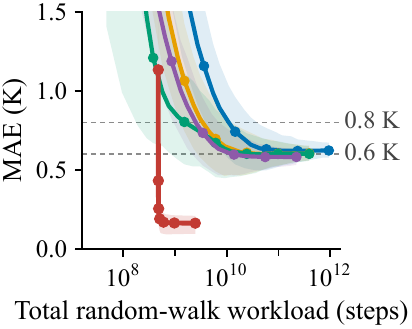}
        }
    \end{minipage}%
    \hfill
    \begin{minipage}[t]{0.16\textwidth}
        \vspace{0pt}
        \centering
        \includegraphics[width=0.90\linewidth]
        {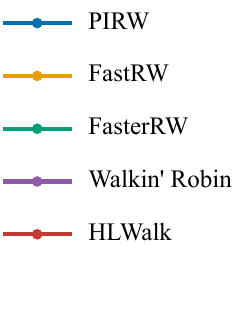}
    \end{minipage}

    \caption{MAE and standard deviation versus total random-walk workload.}
    \label{fig:mae_workload}
\end{figure*}

\section{Experiments}
\label{sec:experiments}

In this section, we evaluate \tool{HLWalk} on three steady-state thermal benchmarks:
Cube, 3DIC, and GPU+HBM, covering multilayer structures
and heterogeneous 2.5D/3D ICs with dense TSVs and large geometric scale disparities.

\subsection{Experimental Settings}
\label{subsec:exp_settings}
For a fair comparison with existing SOTA methods, we
closely follow the setups adopted in prior works\cite{Yang2022:ICCAD,Dong2025:TCAD,Wang2026:DATE_FastRW}. However, due to differences in implementation and
hardware, some details may vary slightly.

\begin{figure}[!t]
    \centering
    \includegraphics[width=0.9\columnwidth]{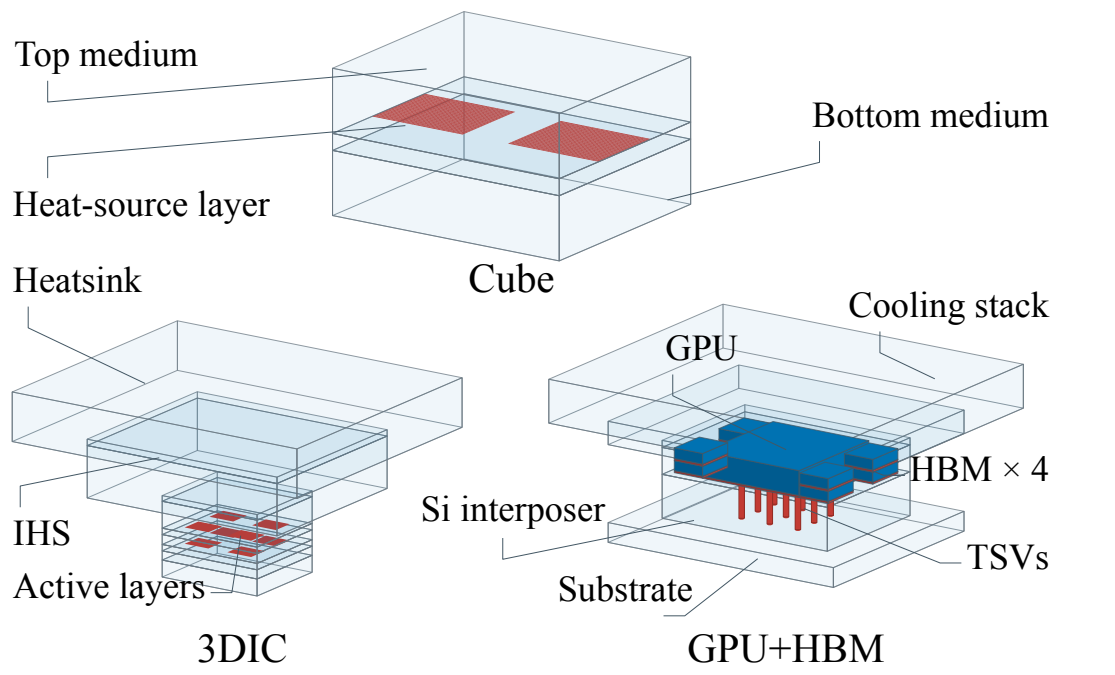}
    \caption{Illustrations of the three thermal benchmarks.}
    \label{fig:test_cases}
\end{figure}

\minisection{Test Cases} Three steady-state benchmarks illustrated in \Cref{fig:test_cases} are considered: Cube, 3DIC, and GPU+HBM.
Cube is a \(20\times20\,\mathrm{mm}^2\) three-layer structure with a
\(100\,\mu\mathrm{m}\) heat-source layer between two
\(500\,\mu\mathrm{m}\) source-free layers, with thermal conductivities
of \(125\) and \(395\,\mathrm{W/(m\cdot K)}\), respectively. Its top
and bottom surfaces use Robin boundaries with
\(h=8700\,\mathrm{W/(m^2\cdot K)}\). The 3DIC case contains 10 material layers and 8 heat sources, with chip/substrate, IHS, and heat-sink footprints of \(20\times20\),
\(40\times40\), and \(60\times60\,\mathrm{mm}^2\), respectively, and
a total power of \(156.48\,\mathrm{W}\). The bottom and top convection
coefficients are \(50\) and \(5000\,\mathrm{W/(m^2\cdot K)}\). GPU+HBM contains one GPU, four dual-layer HBM stacks, 19 heat sources,
and 961 TSVs of \(100\,\mu\mathrm{m}\) radius, with a total power of
\(249\,\mathrm{W}\). The top and bottom convection coefficients are
\(2000\) and \(100\,\mathrm{W/(m^2\cdot K)}\), respectively.

\minisection{Implementation Details}
Each benchmark uses 16 fixed query locations and three random seeds
(42, 137, and 271). COMSOL fine-mesh solutions serve as the golden
reference, while coarse-mesh FEM solutions provide temperature priors\cite{comsol2023}.
For \tool{HLWalk}, the boundary cache uses one path per hemispherical
sampling point with $5\times5$ low-frequency modes for boundary
correction; GPU+HBM uses a higher-order interior modal representation.
We adopt an iso-accuracy setting, selecting the smallest path count that
meets the prescribed MAE target, and all speedups are
measured using end-to-end wall-clock time.

\minisection{Baseline Methods}
We compare \tool{HLWalk} with existing SOTA methods: PIRW\cite{Yang2022:ICCAD}, FastRW\cite{Wang2026:DATE_FastRW}, FasterRW\cite{Wang2026:FasterRW}, and Walkin'
Robin\cite{Miller2024:TOG}. PIRW is a Feynman--Kac-based thermal solver with conventional
path truncation. FastRW accelerates PIRW through truncation-error analysis
and prior-assisted estimation, while FasterRW further improves efficiency
through information-sharing-based post-processing. Walkin' Robin is a
grid-free random-walk solver supporting mixed boundaries.

\subsection{Experiment Results}
\label{subsec:exp_results}

\minisection{Overall Accuracy and Efficiency} We first evaluate the overall accuracy and efficiency of \tool{HLWalk}, focusing on its random-walk workload and iso-accuracy end-to-end performance across the three benchmarks.

\Cref{fig:mae_workload} shows the MAE versus total random-walk workload across the three evaluated benchmarks. HLWalk consistently reaches the high-accuracy regime with substantially lower random-walk workload than the baseline methods. For Cube and 3DIC, where the evaluated methods attain the 0.6-K accuracy target, we compare their computational costs at this common target for a fair comparison. In both cases, HLWalk requires substantially fewer random-walk steps, with the advantage becoming particularly pronounced in the high-accuracy regime. At the plateau operating points on GPU+HBM, \tool{HLWalk} achieves a substantially lower MAE while requiring a much smaller random-walk workload than the baseline methods.

\begin{table}[!t]
\centering
\caption{Iso-accuracy comparison. WinR abbreviates Walkin' Robin.} 
\label{tab:overall_results}
\scriptsize
\setlength{\tabcolsep}{2.0pt}
\resizebox{\columnwidth}{!}{
\begin{tabular}{clrrccc}
\toprule
\tabincell{c}{Case /\\Target}
& Method & $N$ & Steps/path & MAE (K) & Time (s) & Speedup \\
\midrule

\multirow{5}{*}{\tabincell{c}{Cube\\0.6 K}}
& PIRW~\cite{Yang2022:ICCAD}
& 432 & $5.99{\times}10^6$ & 0.584 & 39.27 & 1.00$\times$ \\

& FastRW~\cite{Wang2026:DATE_FastRW}
& 320 & $2.28{\times}10^6$ & 0.483 & 12.10 & 3.25$\times$ \\

& FasterRW~\cite{Wang2026:FasterRW}
& 248 & $2.28{\times}10^6$ & 0.505 & 9.20 & 4.27$\times$ \\

& WinR~\cite{Miller2024:TOG}
& 704 & $9.28{\times}10^4$ & 0.541 & 0.94 & 41.78$\times$ \\

& \tool{HLWalk}
& \textbf{11}
& \textbf{$2.23{\times}10^2$}
& \textbf{0.552}
& \textbf{0.57}
& \textbf{68.89$\times$} \\

\midrule

\multirow{5}{*}{\tabincell{c}{3DIC\\0.6 K}}
& PIRW~\cite{Yang2022:ICCAD}
& 8704 & $2.91{\times}10^5$ & 0.567 & 28.88 & 1.00$\times$ \\

& FastRW~\cite{Wang2026:DATE_FastRW}
& 4864 & $1.42{\times}10^5$ & 0.534 & 10.49 & 2.75$\times$ \\

& FasterRW~\cite{Wang2026:FasterRW}
& 1152
& $1.42{\times}10^5$
& 0.558
& 2.79
& 10.35$\times$ \\

& WinR~\cite{Miller2024:TOG}
& 7680 & $2.41{\times}10^4$ & 0.546 & 3.22 & 8.97$\times$ \\

& \tool{HLWalk}
& \textbf{1122}
& \textbf{$8.26{\times}10^2$}
& \textbf{0.597}
& \textbf{1.23}
& \textbf{23.48$\times$} \\

\midrule

\multirow{5}{*}{\tabincell{c}{GPU+HBM\\Plateau\textsuperscript{a}}}
& PIRW~\cite{Yang2022:ICCAD}
& 2048 & $8.92{\times}10^5$ & 0.660 & 32.33 & -- \\

& FastRW~\cite{Wang2026:DATE_FastRW}
& 1536 & $3.74{\times}10^5$ & 0.650 & 18.15 & -- \\

& FasterRW~\cite{Wang2026:FasterRW}
& 1536 & $3.74{\times}10^5$ & 0.635 & 18.55 & -- \\

& WinR~\cite{Miller2024:TOG}
& 3072 & $2.11{\times}10^5$ & 0.612 & 17.77 & -- \\

& \tool{HLWalk}
& \textbf{768}
& \textbf{$1.90{\times}10^3$}
& \textbf{0.199}
& \textbf{3.34}
& -- \\

\bottomrule
\end{tabular}
}

\vspace{1mm}
\parbox{\columnwidth}{\scriptsize
\textsuperscript{a} On GPU+HBM, the baseline methods fail to reach the 0.6-K target within the evaluated budget. Therefore, we compare the methods at their plateau operating points.
}

\end{table}

\Cref{tab:overall_results} further reports the corresponding
iso-accuracy operating points and end-to-end runtime. On Cube at the
0.6-K target, \tool{HLWalk} achieves an MAE of 0.552~K in 0.57~s,
corresponding to a \textbf{68.89$\times$} speedup over PIRW. On 3DIC at
the same 0.6-K target, it achieves an MAE of 0.597~K in 1.23~s,
yielding a \textbf{23.48$\times$} speedup. On GPU+HBM, at the plateau operating points, \tool{HLWalk} achieves an MAE of 0.199~K in 3.34~s, compared with 0.612--0.660~K and 17.77--32.33~s for the baselines, demonstrating clear advantages in both accuracy and runtime.

\begin{figure}[!t]
    \centering

    \includegraphics[width=0.95\columnwidth]
    {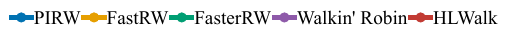}

    \vspace{-1mm}

    \begin{minipage}[t]{0.49\columnwidth}
        \centering
        \includegraphics[width=\linewidth]
        {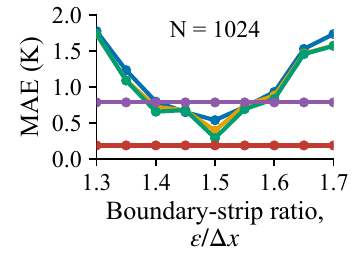}
        \captionsetup{
            font=normalsize,
            justification=raggedright,
            singlelinecheck=false
        }
        \captionof{figure}{Sensitivity to strip ratios $\eps/\Delta x$.}
        \label{fig:rho_sensitivity}
    \end{minipage}%
    \hfill
    \begin{minipage}[t]{0.49\columnwidth}
        \centering
        \includegraphics[width=\linewidth]
        {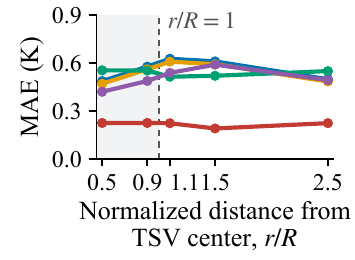}
        \captionsetup{
            font=normalsize,
            justification=raggedright,
            singlelinecheck=false
        }
        \captionof{figure}{Accuracy near TSVs.}
        \label{fig:tsv_accuracy}
    \end{minipage}
\end{figure}

\minisection{Sensitivity to Local-Time Discretization}
We evaluate sensitivity to the normalized boundary-strip width
$\epsilon/\Delta x$ with a fixed path count of $N=1024$. As shown in
\Cref{fig:rho_sensitivity}, PIRW, FastRW, and FasterRW exhibit
noticeable MAE variations as $\epsilon/\Delta x$ changes, reflecting
their sensitivity to local-time discretization. In contrast,
\tool{HLWalk} avoids numerical local-time approximation and is therefore independent of $\epsilon/\Delta x$.

\minisection{Accuracy Around Fine-Scale Structures}
To evaluate local accuracy under strong geometric scale disparities and the advantages of scale-aware random walk, we select six representative TSVs from GPU+HBM to evaluate the MAE at different distances from them. All methods use 8192 paths per query.

As shown in \Cref{fig:tsv_accuracy}, \tool{HLWalk} maintains an
MAE of approximately 0.19--0.23~K over the evaluated distances,
whereas the baseline methods mainly exhibit errors of 0.42--0.63~K.
These results demonstrate that \tool{HLWalk} provides significant advantages over the baseline methods around fine-scale structures by using its scale-aware WOG/WOS strategy.

\subsection{Further Discussions}
\label{subsec:further_discussion}

\minisection{Small-Cache Boundary Acceleration}
We evaluate the effect of small-cache boundary sampling by testing the raw BIE-Random Walk method with only 1 and 16 paths per sampling point to quantify the cost reduction achieved by small-cache sampling alone.

\begin{table}[!t]
\centering
\caption{Effect of small-cache boundary sampling.}
\label{tab:cache_ablation}
\footnotesize
\setlength{\tabcolsep}{3.0pt}

\begin{tabular}{llcccc}
\toprule
Case & Sparse Sampling & N & Steps & Time (s) & Speedup \\
\midrule
\multirow{2}{*}{Cube}
& No  & 16 & $2.77\times10^8$ & 2.68 & 1.00$\times$ \\
& Yes &  1 & $1.74\times10^7$ & \textbf{1.32} & \textbf{2.03$\times$} \\
\midrule
\multirow{2}{*}{3DIC}
& No  & 16 & $1.61\times10^9$ & 8.35 & 1.00$\times$ \\
& Yes &  1 & $1.01\times10^8$ & \textbf{1.94} & \textbf{4.30$\times$} \\
\midrule
\multirow{2}{*}{GPU+HBM}
& No  & 16 & $7.69\times10^9$ & 46.97 & 1.00$\times$ \\
& Yes &  1 & $4.80\times10^8$ & \textbf{6.63} & \textbf{7.08$\times$} \\
\bottomrule
\end{tabular}
\end{table}

As shown in \Cref{tab:cache_ablation}, reducing boundary sampling from 16 paths to 1 lowers the boundary random-walk workload by about 16$\times$ across all benchmarks, yielding approximately \textbf{2.03$\times$}, \textbf{4.30$\times$}, and \textbf{7.08$\times$} speedups on Cube, 3DIC, and GPU+HBM. This confirms that small-cache sampling itself can effectively reduce BIE boundary-reconstruction overhead and the benefit becomes more pronounced as the boundary reconstruction cost increases.

\minisection{Low-Mode Accuracy Enhancement}
We evaluate whether low-mode enhancement can further improve the
accuracy of sparse boundary sampling with limited additional overhead by ablating the low-mode enhancement component.

\begin{table}[!t]
\centering
\caption{Effect of low-mode enhancement.}
\label{tab:accuracy_ablation}
\footnotesize
\setlength{\tabcolsep}{3.0pt}
\begin{tabular}{llcc}
\toprule
Case & Method & MAE (K) & Time (s) \\
\midrule

\multirow{3}{*}{Cube}
& Raw BIE-Random Walk
& 0.432
& 2.68 \\

& \tool{HLWalk} - low-mode
& 1.060
& 1.31 \\

& \textbf{\tool{HLWalk}}
& \textbf{0.325}
& \textbf{1.37} \\

\midrule

\multirow{3}{*}{3DIC}
& Raw BIE-Random Walk
& 0.441
& 8.35 \\

& \tool{HLWalk} - low-mode
& 0.787
& 2.21 \\

& \textbf{\tool{HLWalk}}
& \textbf{0.392}
& \textbf{2.82} \\

\midrule

\multirow{3}{*}{GPU+HBM}
& Raw BIE-Random Walk
& 0.465
& 46.97 \\

& \tool{HLWalk} - low-mode
& 1.198
& 8.19 \\

& \textbf{\tool{HLWalk}}
& \textbf{0.165}
& \textbf{8.57} \\

\bottomrule
\end{tabular}
\end{table}

As shown in \Cref{tab:accuracy_ablation}, small-cache sampling substantially
reduces the boundary-processing cost but introduces noticeable Monte Carlo
error in the estimated boundary field. Low-mode enhancement improves the
boundary estimate by replacing selected low-frequency components with the
deterministic low-mode approximation while retaining the remaining components
from Monte Carlo sampling. This reduces the MAE from 1.060 to 0.325~K on
Cube, 0.787 to 0.392~K on 3DIC, and 1.198 to 0.165~K on GPU+HBM. Meanwhile,
the runtime increases only from 1.31 to 1.37~s, 2.21 to 2.82~s, and 8.19 to
8.57~s, respectively. Moreover, \tool{HLWalk} achieves lower MAE
than Raw BIE-Random Walk on all three benchmarks while still requiring
substantially less runtime. These results show that low-mode enhancement
significantly improves the accuracy of sparse boundary estimation with only
modest additional computational overhead.

\minisection{Residual Decomposition}
We isolate the effect of residual decomposition while keeping the
small-cache configuration and low-mode enhancement unchanged.
Both variants use identical random-walk settings except for the path count.

\begin{table}[!t]
\centering
\caption{Iso-accuracy ablation of residual decomposition.}
\label{tab:residual_ablation}

\resizebox{.8\linewidth}{!}{
\begin{tabular}{llrrr}
\toprule
\tabincell{l}{Case /\\Target}
& Residual
& MAE (K)
& $N$
& Speedup \\
\midrule

\multirow{2}{*}{\tabincell{l}{Cube\\0.6 K}}
& No  & 0.580 & 13 & 1.00$\times$ \\
& Yes & 0.552 & \textbf{11} & \textbf{1.18$\times$} \\
\midrule

\multirow{2}{*}{\tabincell{l}{3DIC\\0.6 K}}
& No  & 0.545 & 3136 & 1.00$\times$ \\
& Yes & 0.597 & \textbf{1122} & \textbf{2.80$\times$} \\
\midrule

\multirow{2}{*}{\tabincell{l}{GPU+HBM\\0.8 K}}
& No  & 0.645 & 42 & 1.00$\times$ \\
& Yes & 0.711 & \textbf{19} & \textbf{2.21$\times$} \\
\bottomrule
\end{tabular}}
\end{table}

As shown in \Cref{tab:residual_ablation}, residual decomposition reduces the path count required to meet the target accuracy on the heterogeneous benchmarks by removing the smooth large-scale
temperature component before stochastic sampling. The sampling speedups reach \textbf{1.18}$\times$, \textbf{2.80}$\times$, and \textbf{2.21}$\times$ on Cube, 3DIC, and GPU+HBM, respectively. This indicates that residual decomposition reduces the random-walk sampling burden for heterogeneous multiscale structures.

%% file: docs/conclusion.tex
\section{Conclusion}
This paper introduced \tool{HLWalk}, a random-walk solver for thermal
analysis of heterogeneous 2.5D/3D ICs. \tool{HLWalk} eliminates
local-time boundary discretization error and reduces boundary-processing
cost through modal-split acceleration, significantly decreasing the boundary
random-walk workload while improving
accuracy. Its scale-aware WOS/WOG transport further maintains high
accuracy under large geometric scale disparities. Overall,
\tool{HLWalk} achieves up to 23$\times$ end-to-end wall-clock speedup
on heterogeneous 2.5D/3D ICs. These results highlight \tool{HLWalk} as an efficient solver for heterogeneous IC thermal analysis, with future extensions to transient thermal simulation.